\documentclass[seceq]{ptptex}

\usepackage{graphicx}
\begin{document}

\title{Sequence of multipolar transitions: Scenarios for URu$_2$Si$_2$}


\author{Patrik Fazekas$^1$, Annam\'aria Kiss$^2$, and Katalin Radn\'oczi$^1$ \\
$^1$Research Institute for Solid State Physics and Optics,\\
Budapest 114, P.O.B. 49, H-1525 Hungary\\
$^2$Department of Physics, Tohoku University, Sendai, Japan.}

\maketitle

\begin{abstract}
$d$- and $f$-shells support a large number of local degrees of freedom: dipoles,
quadrupoles, octupoles,
hexadecapoles, etc. Usually, the ordering of any multipole component leaves the
system sufficiently symmetrical to allow a second symmetry breaking transition.
{\sl Assuming that a second continuous phase
transition occurs}, we classify the possibilities.  We construct the symmetry group
of the first ordered phase, and then re-classify the order parameters in the
new symmetry. While this is straightforward for
dipole or quadrupole order, it is less familiar for octupole order.

We give a group theoretical analysis, and some illustrative mean field
calculations, for the hypothetical case when a second ordering transition
modifies the primary ${\cal T}_{xyz}$ octupolar
ordering in a tetragonal system like URu$_2$Si$_2$. If quadrupoles appear
in the second phase transition, they must be accompanied by a time-reversal-odd
multipole as an induced order parameter. For
${\cal O}_{xy}$, ${\cal O}_{zx}$, or ${\cal O}_{yz}$ quadrupoles, this
would be one of the components of {\bf J}, which should be easy either to check
or to rule out. However, a pre-existing octupolar
symmetry can also be broken by a transition to a new octupole--hexadecapole
order,   or by a combination of ${\cal O}_2^2$ quadrupole and triakontadipole order.

It is interesting to notice that if  recent NQR results\cite{takagi}
on URu$_2$Si$_2$ are interpreted as a hint that  the onset of octupolar hidden
order  at $T_0=17{\rm K}$
is followed by quadrupolar ordering at $T^*\approx 13.5{\rm K}$, this sequence
of events may fit several
of the scenarios found in our general classification scheme. However, we have
to await further evidence
showing that the NQR anomalies at $T^*\approx 13.5{\rm K}$ are associated with
an equilibrium phase transition.
\end{abstract}

\section{Introduction}

There is increasing interest in orbital ordering phenomena, and their  relationship to
magnetism\cite{ORBITAL,ASR2002,erdosreview}. Many of these phenomena can be at least partially
understood in terms of localized electron models. This is clearly  justified for Mott-localized
$d$-electrons. Some  $f$-electron systems are semiconductors (like NpO$_2$), but more often,
we wish to describe multipole ordering in metals like rare-earth-filled skutterudites or
URu$_2$Si$_2$. In the standard model\cite{standard} of the majority of rare earth elements the
$f$-shell is localized, and we may invoke a similar feature for the systems of interest to us.
It can be argued that itineracy  and multipolar ordering  are complementary features, and
that they may be manifest in different phases of the same $f$-electron system. In this case,
the localized description is acceptable for the ordered phases, even though we know that it
could not be extended to the whole phase diagram.

The common starting point is the existence of an $N$-dimensional local Hilbert space
$|n\rangle$, which allows the definition of $N^2$ local operators $|n\rangle\langle m|$
 (of these, $N^2-1$ are non-trivial, and are also called local order parameters).
 $N(N+1)/2$ operators have symmetrical, and $N(N-1)/2$ operators have antisymmetrical
 character. They can be chosen as
 \begin{equation}
 X_{mn}^s = \frac{1}{2} ( |m\rangle\langle n| + |n\rangle\langle m|)
  \label{eq:sn5}
 \end{equation}
for symmetrical and
 \begin{equation}
 X_{mn}^a = \frac{1}{2i} ( |m\rangle\langle n| - |n\rangle\langle m|)
  \label{eq:sn6}
 \end{equation}
 for antisymmetrical  operators where $i$ was inserted to ensure  hermiticity.

 There are two canonical cases: the basis  may consist of
\renewcommand{\labelenumi}{\Roman{enumi}.}
\begin{enumerate}
\item{N/2 time-reversed pairs (the  N-fold degeneracy arises as  2${\times}$(N/2),
\newline (Kramers degeneracy)${\times}$(non-Kramers degeneracy)}
\item{real basis states each of which is time reversal invariant (the N-fold degeneracy
 is purely non-Kramers degeneracy)}
\end{enumerate}

{\bf Case I.} The local Hilbert space consists of $(N/2)$ time-reversed pairs. There
are $N(N-1)/2$ time-reversal-even order parameters (this includes the trivial ${\hat 1}$),
and $N(N+1)/2$ time-reversal-odd order parameters.

$N=4$ is realized by the $\Gamma_8$ representation of the cubic double group, the irrep of
the ground state level for either the $4f^1$ compound CeB$_6$ (Refs. \cite{Ohkawa,shiina97}),
or the   $5f^3$ compound NpO$_2$ (Ref. \cite{fournier}). Six operators: ${\hat 1}$ and the
five quadrupoles are time-reversal-even, while ten order parameters: the three dipoles
and seven magnetic octupoles, change sign under time reversal. A general form of intersite
interaction is a sum over the  quadratic invariants, for a cubic system with six independent
coupling constants (one dipolar, two quadrupolar, three octupolar). It was, however, soon
realized\cite{Ohkawa} that the consideration of the SU(4) symmetrical model (with
all couplings set equal) should be enlightening. Later research showed that
CeB$_6$ can be regarded as a "nearly SU(4)-symmetrical" system\cite{Kusu2001,shiina2002}.

{\bf Case II.} If all basis states can be chosen real, the $X_{m,n}^s$ are real operators
and the $X_{m,n}^a$ are imaginary operators. Choosing one of the real operators as the
projection onto the entire local Hilbert space ${\hat {\boldmath 1}}=\sum_m  |m\rangle\langle m| $,
 and orthogonalizing the remaining $N-1$ diagonal operators, we are left with $N(N+1)/2 - 1$
 time-reversal invariant local order parameters.
$N(N-1)/2$ local order parameters  (the imaginary operators $X_{m,n}^a$) change sign under time reversal.

This case is realized for non-Kramers ions which have an even number of electrons. We are
interested in the $5f^2$ (U$^{4+}$), and $4f^2$ (Pr$^{3+}$) configurations as they appear
in URu$_2$Si$_2$ and PrFe$_4$P$_{12}$, resp. It is always debatable which $N$ to choose.
Crystal field doublets ($N=2$) have been suggested for both systems but it is unlikely
for PrFe$_4$P$_{12}$, and at least not widely accepted for URu$_2$Si$_2$. In fact, for
both systems we settle for a local Hilbert space which is composed of the bases of
several irreps. For PrFe$_4$P$_{12}$  the quasi-quartet\cite{kifa}
$\Gamma_1+\Gamma_4$ seems to work. Under tetrahedral symmetry, the nine time-reversal-even
order parameters are a singlet,
five  quadrupoles, and three hexadecapoles, while the six time-reversal-odd order
parameters are three dipoles and three octupoles\cite{shiina}.

\subsection{The nature of the higher multipoles}

While dipoles and quadrupoles are well-known from decades of research experience, it is only recently that
octupoles  were seriously considered, and higher multipoles are virtually never discussed. It is worth
forming a picturial idea of octopule-, hexadecapole-, and triakontadipole-carrying shells.

The $J=4$ Hund's rule two-electron states of U$^{4+}=5f^2$ ionic cores are rather complicated objects,
superposed of a number of $l_js_j$ components by the Clebsch-Gordan coefficients.  However,  there is
strong similarity between the  multipoles composed of two $f$-orbitals and two spins via a number of
projections, and simplified multipoles built of $l=4$ orbitals (atomic $g$-states).  We derive the charge
cloud shapes and current distribution patterns  from acting with the analogous purely orbital (e.g.,
${\overline{J_z(J_x^2-J_y^2)}}\rightarrow{\overline{l_z(l_x^2-l_y^2)}}$) operators on fictitious
atomic states (for our purposes,  fourth-order spherical harmonics   $Y^4_m(\vartheta,\phi)$ for
$g$-states). In this Section, we do not discuss  crystal field effects.

We diagonalize multipole operators  in the 9-dimensional $l=4$ space, and base our pictures on the
state with the (in absolute value) largest eigenvalue. This can be interpreted as finding the ground
state of an effective field which orders the multipole. We mention that free-ion multipole eigenstates
are not necessarily the same that we find within the restricted Hilbert spaces of crystal field
problems. This will be illustrated by comparing Fig.~\ref{fig:octl=4} and
Fig.~\ref{fig:t1t3}. Furthermore, smaller Hilbert spaces may not give a representation of certain
multipoles at all.

For odd-rank (magnetic) multipoles, the eigenstates appear in time-reversed pairs, similar to
the $l_z=\pm m$ dipole eigenstates, and the multipole spectrum is symmetrical about 0. For
even-rank (electric) multipoles, each eigenstate can be chosen time-reversal invariant.

\subsubsection{Octupoles}

First, let us consider the octupole
\begin{equation}
{\cal T}_{xyz} \rightarrow \frac{1}{6}{\overline{l_xl_yl_z}} = \frac{-i}{4}\left(
l_+l_zl_- -l_-l_zl_-\right)
\label{eq:octu}
\end{equation}

The simplest octupolar states are realized in the $l=2$ subspace of $e_g$ functions
\begin{equation}
\Phi_{\pm}(l=2) = \frac{1}{\sqrt{2}}\left(|3z^2-r^2\rangle \pm  i | x^2-y^2\rangle \right)
\label{eq:octul2}
\end{equation}
$\Phi_{\pm}(l=2)$ form a time-reversed pair.

\begin{figure}[ht]
\vspace{-0.2cm}\centerline{\includegraphics[height=5 cm] {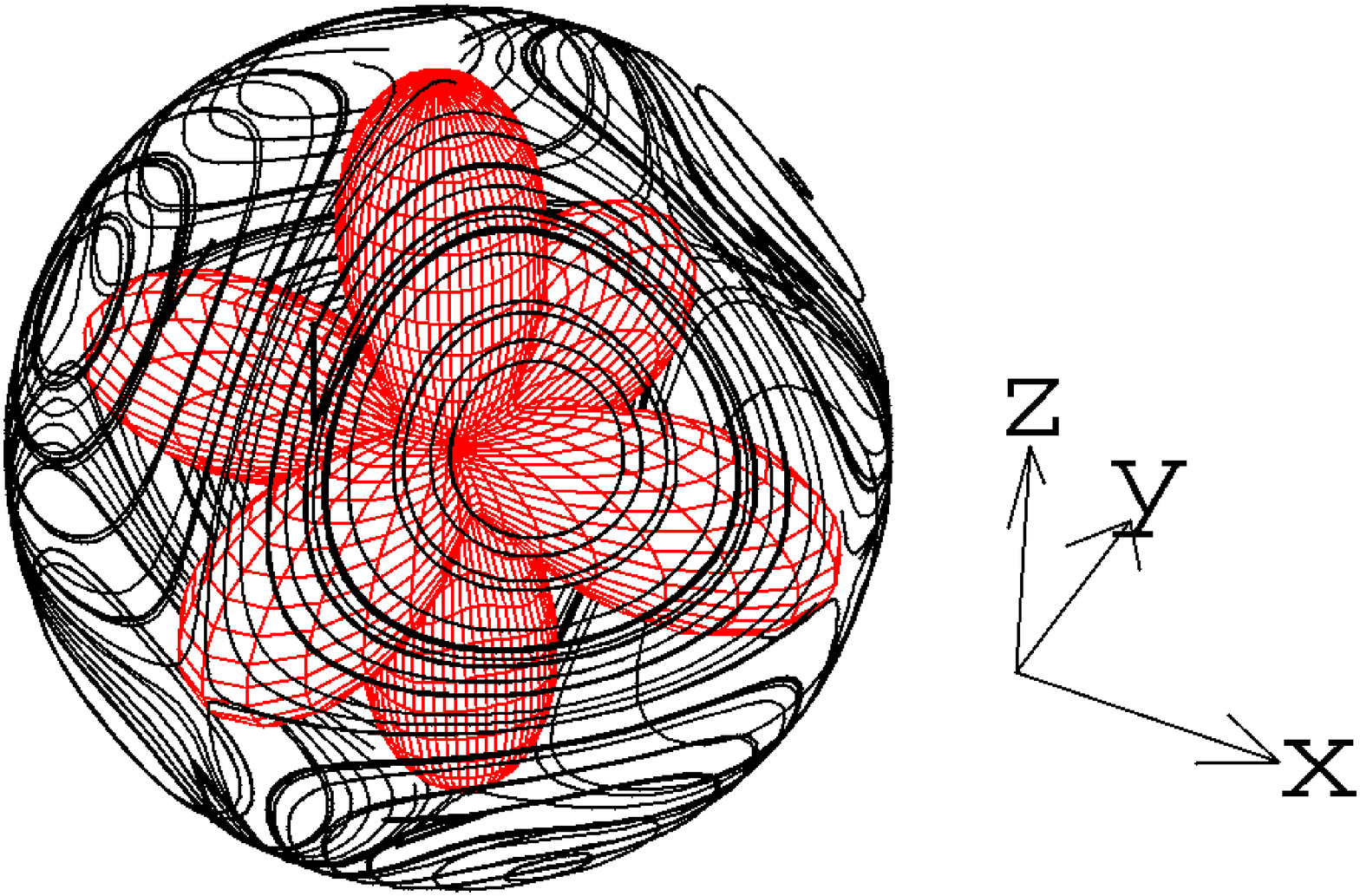}\hspace{0cm}
\includegraphics[height=5cm]{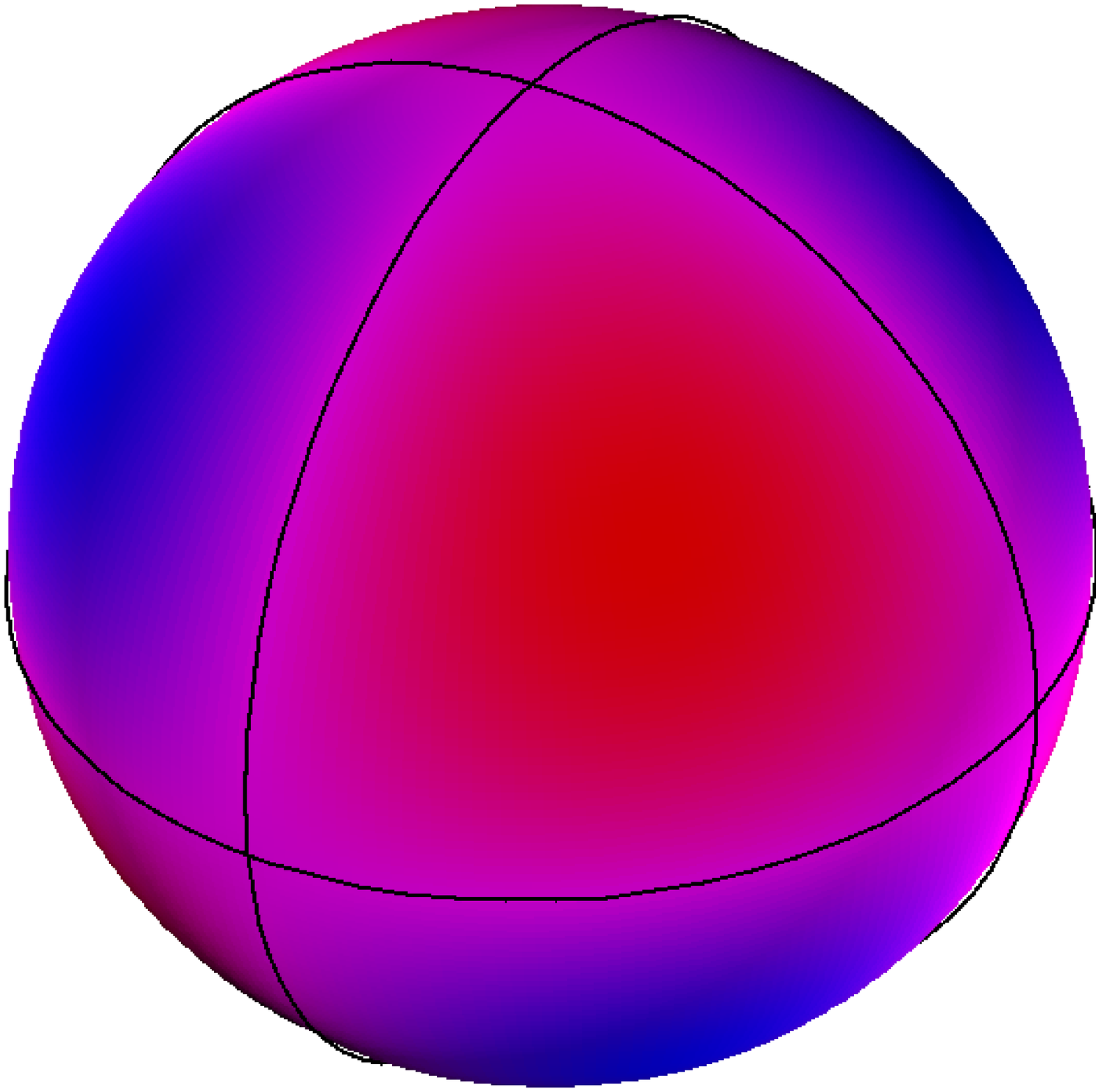}}  \vspace{-0cm}
\caption{Left: Charge distribution (lobes) and current pattern (flow lines) in
the ${\cal T}_{xyz}$ octupole eigenstate $\Phi_{+}$ in the $l=2$ Hilbert space.
 Right: the normal component of the magnetic field is pointing inwards (outwards)
 in the light grey (dark grey) areas.} \label{fig:octupole} \end{figure}

\begin{figure}[ht]
\vspace{-.2cm}\centerline{\includegraphics[height=6cm] {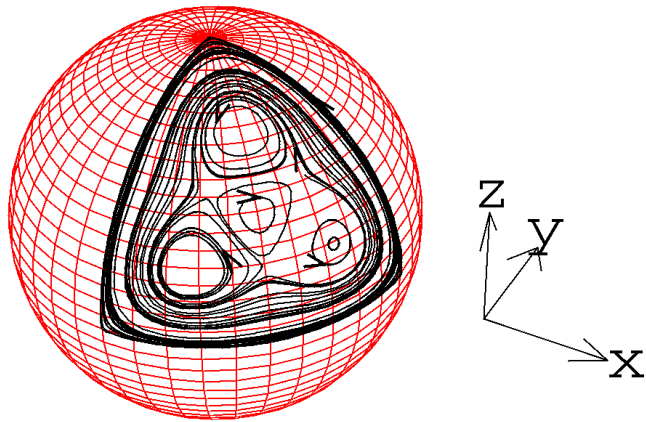}}
\vspace{0cm}\caption{Current distribution in the ${\cal T}_{xyz}$ octupole
ground state (\protect\ref{eq:octu2}) in the $l=4$ Hilbert space. Only one octant
is shown.}
\label{fig:octl=4}
\end{figure}

The charge and the current distribution of $\Phi_+(l=2)$ is shown in Fig.~\ref{fig:octupole}.
The object got its name from the eight magnetic poles: of the eight current eddies we
see in the figure\footnote{Two  general remarks about Figures
 ~\ref{fig:octupole}--\ref{fig:t1t3}: charge density angular dependences are shown,
  i.e., the lobe shapes do not include the radial fall-off of the atomic wave functions.
Flow lines are to indicate the sense of circulation of the current but
(in contrast to the textbook interpretation) the density of flow lines is not
associated with higher current density, but is rather arbitrary. Flow lines were
calculated by solving the differential equation for the tangential curves for
the calculated vector field of currents, and initial values were randomly generated.
(We found that the direct representation of the vectors would give unattractive figures).},
four belong to magnetic field lines entering, and four to those leaving the surface.
Note that the magnetic field pattern measured\cite{musroct2000} by $\mu$SR in NpO$_2$ bears
an overall similarity to what we expect in the neighborhood of an octupole-moment
bearing shell. For the time-reversed partner  $\Phi_-(l=2)$ we would find the same
charge cloud with reversed currents.

The concept of octupolar ordering was pioneered by Korovin and Kudinov\cite{korovin}
who envisaged an antiferro-octupole pattern of the $e_g$ states (\ref{eq:octul2})
resulting from spin-orbital exchange in Mott insulators. A similar possibility
arises in the $E$ doublets of trigonal compounds\cite{vernay} where, however,
the octupole moment is  mixed with the  orbital moment $l_z$. In any case,
antiferro-orbital order is likely to be combined with spin ferromagnetism\cite{TS,Khm,mae}.
Metallic phases, including the itinerant octupolar phase, were investigated for the
$e_g$ band Hubbard model by Takahashi and Shiba\cite{TS}.

Currently known realizations of octupolar order appear in $f$-electron systems. Field-induced
octupoles play a role in understanding the phase diagram of CeB$_6$ (Ref.~\cite{shiina97}).
Kusunose and Kuramoto\cite{KK1} called attention to the fact that octupole moments
are ideally suited for the role of "hidden" primary order parameters. A  detailed
study by Kubo and Kuramoto\cite{KK} makes a convincing case that the "Phase IV" of
Ce$_{1-x}$La$_x$B$_6$ is an antiferro-octupolar phase. At about the same, it became
accepted that the long-standing mystery of the nature of the 25K transition of
NpO$_2$ is solved by identifying it with the triple-{\bf q} ordering of $\Gamma_{5u}$
octupoles\cite{npo202,kf2003,hotta}. A symmetry analysis was successful in
identifying the unique octupolar signature in NMR spectra on NpO$_2$, and Ce$_{1-x}$La$_x$B$_6$
(Ref.~\cite{sakai05}. The phenomenology of the two systems shows certain similarities,
as it is to be expected since, e.g., the relationship between the anomalies of the
linear, and the third-order, susceptibilities follow from general thermodynamic
reasoning\cite{kf2003,sakak}. The same should be true of URu$_2$Si$_2$ whos
hidden order, we argued\cite{kf}, is also of octupolar nature.

Fig.~\ref{fig:octupole} illustrates the symmetry of the (uniform) octupolar ground
state with $\langle{\cal T}_{xyz}\rangle$ as the order parameter. We leave to
Sec.~\ref{sec:octupole} the construction of an octupolar symmetry group, which will
be carried out for a combination of tetragonal crystal field and octupolar effective
field, the case relevant for URu$_2$Si$_2$. Here we use  Fig.~\ref{fig:octupole}
to visualize the hybrid nature of some of the symmetry elements. The charge distribution
is highly symmetrical (octahedral). However, the true symmetry is that of the
magnetic field pattern, so in purely geometrical terms it is tetrahedral. It is obvious,
however, that ${\cal C}_4$ rotations are allowed if they are combined with time
reversal ${\cal T}$ (which reverses the currents and the field). Including elements
like ${\cal C}_4{\cal T}$, a higher non-unitary symmetry group  could  be derived.

Figure.~\ref{fig:octupole} shows why octupolar order can be so well "hidden".
The charge distribution (which influences directional bonding, i.e., the structure
of the crystal) is cubic, hiding the fact that if current directions are considered,
the pure ${\cal C}_4$ axes are not symmetry elements. Similar considerations arose
when it was asked whether the  pseudocubic phase of Sr-doped LaMnO$_3$ is not,
in fact, octupolar ordered\cite{TS}.

One might have thought that the character of the octupoles is the same  whatever
Hilbert space we represent them on, but there are interesting nuances here.

Within the  $l=4$ shell (which  mimicks\footnote{However, a calculation of the
currents for realistic two-electron states would be desirable.} URu$_2$Si$_2$'s
$5f^2\rightarrow J=4$ multiplet), ${\cal T}_{xyz}$ has the minimal eigenvalue
$-6\sqrt{3}$, corresponding to the eigenstate (expressed in the $|l_z\rangle$ basis)
\begin{equation}
\Phi^{\rm oct}_0=\sqrt{\frac{7}{50}}(|+4\rangle + |-4\rangle ) - 2i\sqrt{\frac{3}{50}}
 (|+2\rangle + |-2\rangle ) -\sqrt{\frac{1}{5}}
|0\rangle
\label{eq:octu2}
\end{equation}
(the maximal eigenvalue  $6\sqrt{3}$ corresponds to the time-reversed of
$\Phi^{\rm oct}_0$). The current distribution for $\Phi^{\rm oct}_0$ is shown in
Fig.~\ref{fig:octl=4}. The overall pattern of eight alternating vortices is the same as
in Fig.~\ref{fig:octupole}. However, within each of the major eddies four new sub-eddies
appear: three (arranged like the petals of a flower) rotating in the sense of the
eddy as a whole, and a little central eddy counter-rotating. The symmetry of the
current distribution in Fig.~\ref{fig:octl=4} is the same as that in Fig.~\ref{fig:octupole},
but there are differences in the magnetic field pattern. While in Fig.~\ref{fig:octupole}
the magnetic field is maximum in the center of an octant, and has the same sign
everywhere within the octant, in the $l=4$ solution (Fig.~\ref{fig:octl=4}) the field
changes sign in a small central region of the octant and appears smaller than in the
"petals". On the whole, the magnetic field of the octupolar currents is weaker for
$l=4$ states ($f^2$ configurations) than for the simplest $l=2$   solution. This
may account for the weakness of the internal fields in URu$_2$Si$_2$. It should be
noted, though,  that the current distribution shown in Fig.~\ref{fig:octl=4} was
obtained for a free ion subject to the octupolar effective field only. The situation
changes if crystal field effects are included (Sec.~\ref{sec:octupole}).

It would be of obvious interest to calculate internal field distributions for octupoles,
and maybe also for triakontadipoles, for these predictions could be checked against
$\mu$SR and NMR observations. We note that Kubo and Kuramoto\cite{KK} discussed the
situation for the two well-established octupolar systems NpO$_2$ and Ce$_{1-x}$La$_x$B$_6$.
for which 500G and 40G, respectively, are measured; both are in excess of the theoretical
estimate. The measured internal field in URu$_2$Si$_2$ is much weaker: $^{29}$Si-NMR
linewidth gives $\le 12$G (Ref.\cite{bernal}), while at the muon stopping sites, the
internal field is at most 1--2G (Ref.\cite{musr}).

 \subsubsection{Hexadecapoles}

\begin{figure}[ht]
\vspace{-.2cm}\centerline{\includegraphics[height=6.5cm] {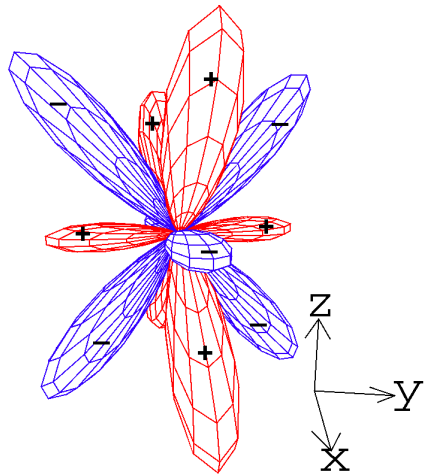}}
\vspace{-.1cm}\caption{Charge distribution in the electric  hexadecapole eigenstate
(\protect\ref{eq:hex}). The sign of the lobes of the wave function is also shown.}
\label{fig:hexa}
\end{figure}

The hexadecapole ${\overline{l_xl_y(l_x^2-l_y^2)}}$ has the largest eigenvalue
for the eigenstate
\begin{equation}
\Phi_{\rm hex}=\frac{1}{\sqrt{2}}(|2\rangle -i|-2\rangle)\, .
\label{eq:hex}
\end{equation}
In spite of the appearance of $i$, $\Phi_{\rm hex}$ is time reversal invariant.
It can be chosen real and its positive and negative lobes can be identified
(Fig.~\ref{fig:hexa}). There is a general similarity to quadrupolar eigenstates (which
also have positive and negative lobes), only the number of lobes is higher. We do
not construct the symmetry group of the hexadecapole effective field but we notice
that, like the octupolar symmetry group, it would have composite symmetry elements:
purely geometrical transformations combined with "lobe sign reversal".

Hexadecapoles as order parameters were discussed within the tetrahedral symmetry
classification\cite{shiina}, valid for Pr-filled skutterudites. We are not aware
of the existence of primary hexadecapole order in any system.

\subsubsection{Triakontadipoles}

\begin{figure}[ht]
 \vspace{0cm}\centerline{\includegraphics[height=5.5 cm] {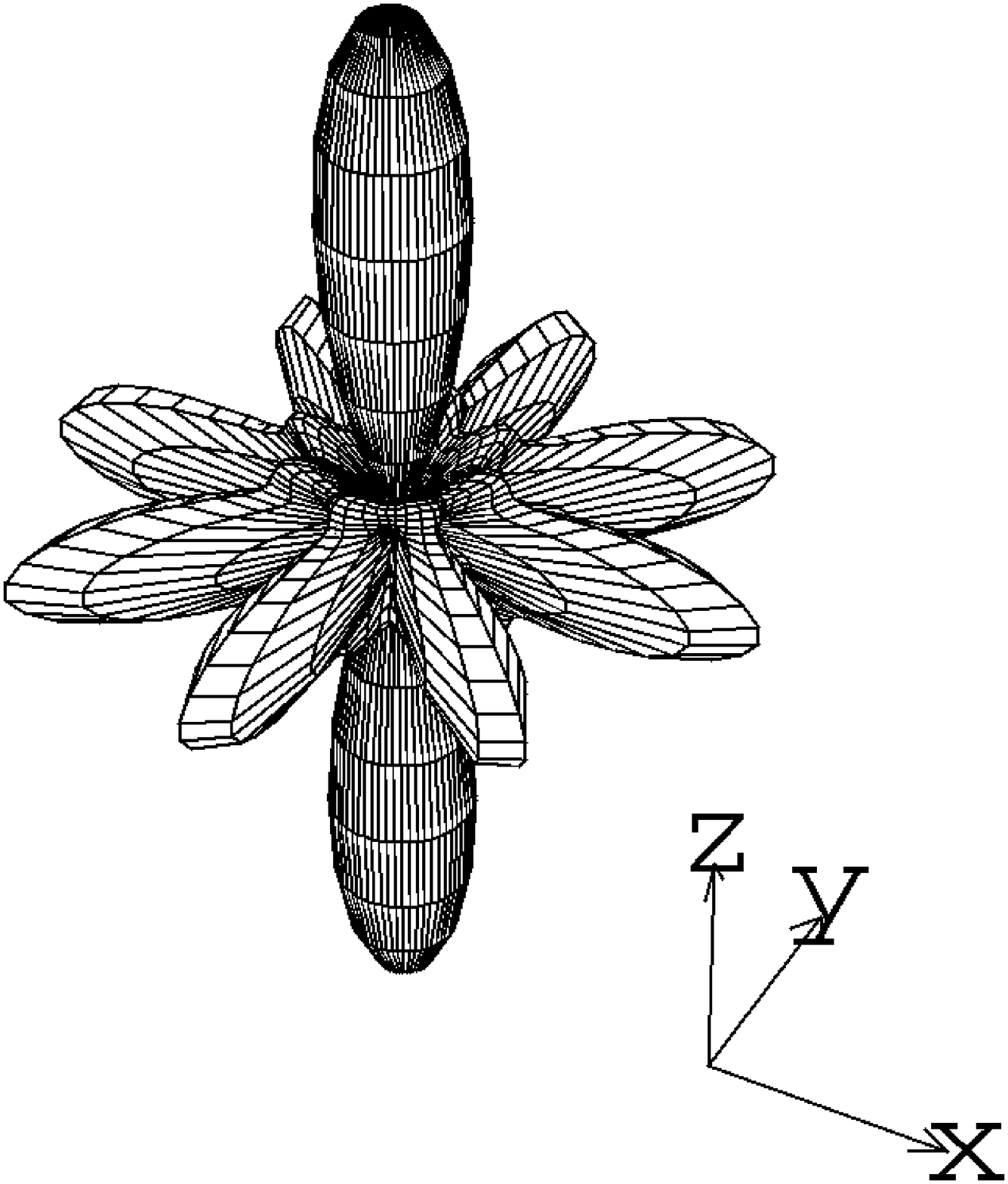}\hspace{0cm}\includegraphics[height=5.5 cm] {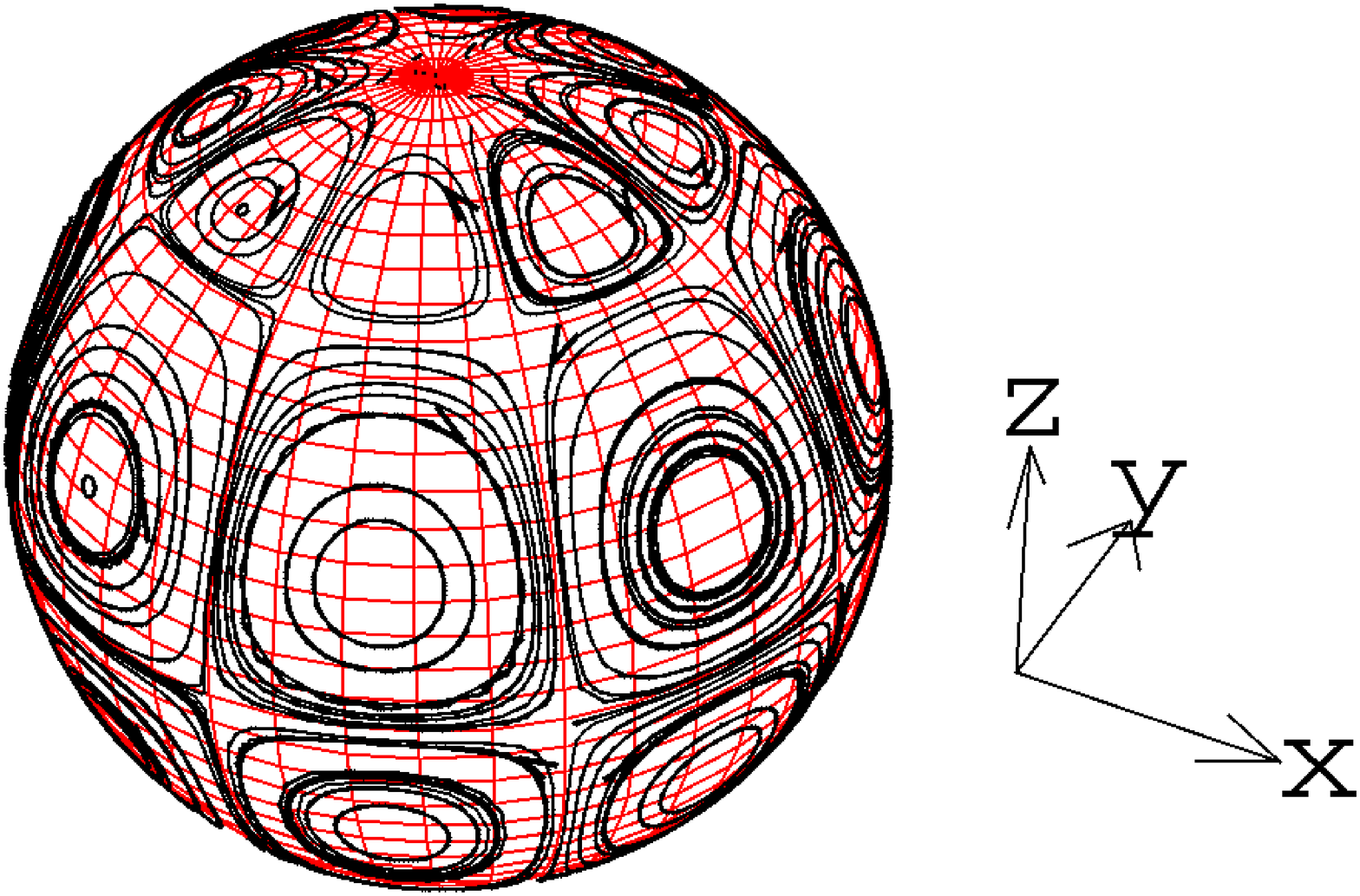}}  \vspace{0cm}\caption{The charge (left) and the current (right) distribution in the magnetic triakontadipole eigenstate
 (\protect\ref{eq:tria}).} \label{fig:triacur}
 \end{figure}

There are 11 triakontadipoles but we consider only  ${\overline{l_xl_yl_z(l_x^2-l_y^2)}}$
which appears as the lowest-rank
$A_{1u}$ multipole in the tetragonal classification (Table~\ref{tab:tetr}).  Its ground
state is
\begin{equation}
\Phi_{\rm tria}=\frac{1}{2}\left( |4\rangle +|-4\rangle +i\sqrt{2}|0\rangle\right)\, .
\label{eq:tria}
\end{equation}
with eigenvalue $6\sqrt{35}$.  As shown in Fig.~\ref{fig:triacur}, there are 32
cells of alternatingly flowing current, so the associated magnetic field is more
short-ranged than in the case of an octupole (Fig.~\ref{fig:octupole}). As in the case
of octupoles, the symmetry of the charge distribution is higher than that of the current
distribution: the charge cloud has  a ${\cal C}_8$ axis which is reduced to ${\cal C}_4$
for the currents. However, the combination of the $\pi/4$ rotation with time reversal
${\cal C}_8{\cal T}$ is a symmetry operation. It would be interesting to meet
triakontadipolar order in nature.

\section{Recent experimental developments on URu$_2$Si$_2$}

The magnetic behavior and phase diagram of the intermetallic compound URu$_2$Si$_2$
have been the focus of attention for over two decades\cite{hiebl}. Specific heat
measurements \cite{schlabitz} show that electronic entropy of $O(\ln{2}){\cdot}k_{\rm B}$
is released by the time the
temperature reaches 30K, and a sizeable fraction of it is associated with the
$\lambda$-anomaly at $T_0\approx 17{\rm K}$. Thus it is not unjustified to think of the
phase transition as the full-scale ordering of a localized degree of freedom, but the
nature of the order parameter remains hidden. It is obviously not the
tiny ($\sim 0.03{\mu}_{\rm B}$) antiferromagnetic moment which is observed by neutron
scattering \cite{broholm}.

Small (either static or slowly fluctuating) moments have long been held to be an
attribute of heavy fermion systems on the borderline between localized and itinerant
$f$-electron phases. The specific heat value would allow to classify  URu$_2$Si$_2$ as
a "light heavy fermion system", raising the question whether its mysterious properties
may be related to  an exotic itinerant phase of strongly correlated $f$-electrons.
Orbital magnetism due to plaquette currents\cite{chandra}, unconventional density
waves\cite{virosztek,zhito}, and Pomeranchuk instability leading to a nematic
state\cite{varma} are in this category.

The RKKY interaction mediates a variety of multipole-multipole interactions between
the $f$-shells\cite{coqblin}. However, the ordering may be foiled by the collective
Kondo effect: the formation of a heavy Fermi sea with a large (Luttinger) Fermi
surface\cite{shifa} may swallow up the localized moments. The Kondo-to-RKKY
transition has been extensively studied for the case when the relevant local
degrees of freedom are spin dipoles\cite{doniach} but work on the general
multipolar problem  has only just started\cite{otsuki}.  It is an intriguing
possibility that the 17K phase transition  in URu$_2$Si$_2$ coincides with the
itinerant-to-localized transition of the $f$-electrons\cite{kuramoto97}.  This
would lend credence to describing the $T<T_0$ order in terms of a localized
$f$-electron model even if a more satisfactory description will have to encompass
itinerant aspects\cite{as in}.

It has long been known\cite{walker,agterberg} that the symmetry classification  of
local order parameters  in tetragonal crystal fields follows Table~\ref{tab:tetr}.
For each of the irreps, only the lowest-rank multipole is listed. The result can
be viewed as arising from the tetragonal splitting of the operator irreps of the
cubic classification scheme\cite{shiina97}.

\begin{table}[ht]\caption{Symmetry classification of the local order 
parameters for ${\bf B} =0$ (${\cal D}_{4h}$ notations \protect\cite{inui}, overline
means symmetrization). $g$ ($u$) belongs to even (odd) parity under time reversal.}
\label{tab:tetr}
\centering
\begin{tabular}{|c|c||c|c|}
\hline \hline sym ($g$)& operator & sym ($u$) & operator\\[1mm]
\hline
$A_{1g} $ &  ${\cal E}$ & $A_{1u} $&$ {\overline{J_x J_y J_z (J_x^2 - J_y^2)}} $\\[1mm]
$A_{2g}$ & ${\cal H}_1={\overline{J_x J_y  (J_x^2 - J_y^2)}} $& $A_{2u}$ & $J_z$ \\
$B_{1g} $ & ${\cal O}_2^2$ &  $B_{1u} $ &  ${\cal T}_{xyz}= {\overline{J_x J_y J_z }}$\\
$B_{2g}$ & ${\cal O}_{xy}={\overline{J_x J_y }}$ & $B_{2u}$ &
${\cal T}^{\beta}_z={\overline{J_z(J_x^2-J_y^2)}}$\\
$E_g$ & $\{ {\cal O}_{xz}, {\cal O}_{yz} \}$ & $E_u$ & $\{ J_x, J_y \}$  \\
\hline
\end{tabular}
\end{table}

Progress with uncovering the true nature of the hidden order  of URu$_2$Si$_2$
has been particularly slow because of a seemingly extrinsic property of almost
 all of the samples: they show a kind of micro-antiferromagnetism with moments
$m\sim 0.02-0.04{\mu_{\rm B}}\parallel{\hat z}$ with the simple alternating
order ${\bf Q}=(0,0,1)$\cite{walker}. However, evidence from microscopic
measurements\cite{matsuda,segreg,amitsuka,musr} shows that the nominally small
moment should be understood as a relatively large moment at a minority of the sites.

The relationship between the hidden order parameter $\psi$ and the antiferromagnetic
moment $m$ has been, and still remains\cite{bourdarot}, a matter of debate.
There are two basic possibilities\cite{shah}: A)  hidden order and antiferromagnetism
share the same symmetry\cite{thesame}, or B) they are of different symmetries.
In Case A), the relative amplitude of $m$ and $\psi$ can be tuned continuously
(e.g., by pressure), and there is no sharp distinction between the two orders.
In Case B), the two orders are incompatible, and there must be a first order
transition from the $\psi\ne 0$, $m=0$ phase to the $\psi=0$, $m\ne 0$ phase.
In the present discussion (as in Ref.\cite{kf}) we take the conclusion drawn
from  recent $\mu$SR experiments\cite{musr} as our starting point: there is
a first-order transition, the symmetry of $\psi$ is different from that of
$m$, and therefore the hidden order parameter must be sought from among other
entries in Table~\ref{tab:tetr}.

As far as gross features like linear and non-linear\cite{ramirez} susceptibility,
specific heat, etc. are concerned, with a little adjustment of the crystal field
level scheme, $B_{1g}$ and $B_{2g}$ quadrupolar \cite{santini1}, and  $B_{1u}$
and $B_{2u}$ octupolar \cite{kf} models do about equally well. Evidence beyond
this simple range of experiments has to be invoked to choose between the
quadrupolar and octupolar scenarios.

We cite two crucial (but as yet unpublished) experiments to argue that the hidden
order is not quadrupolar. First, there is a remarkable mechanical--magnetic
cross--effect (at $T<17$K): in the presence of uniaxial stress applied perpendicularly
to the tetragonal main axis, large-moment
antiferromagnetism with the simple pattern described above becomes visible for
neutron magnetic Bragg scattering \cite{yokoyama}. This is unambiguous proof that
the background order breaks time reversal invariance, and is thus certainly not
quadrupolar; the simplest remaining choice is octupolar order.

The second evidence is coming from recent NQR measurements by Saitoh et al
\cite{takagi}. The temperature dependence of the electric field gradients was
followed carefully from relatively high temperatures (70K) to well below
$T_0=17$K. $V_{zz}$ changes all the time, reflecting the $T$-dependence of
the tetragonal crystal field component ${\cal O}_2^0$. Saitoh et al's findings
can be formulated in  two statements: (a) Neither of the field gradients shows
anomalies at  $T_0=17$K, the onset temperature of hidden order, so the hidden
order cannot be quadrupolar. (b) There is an anomaly in the NQR signal at
$T^*\approx 13.5$K, so something happens to the quadrupoles there.  A possibility
is that $T^*$ is the ordering transition of quadrupoles; then it has to be a second
HO transition following the first one at $T_0$. We emphasize that a lot more
experimental evidence (especially specific heat, susceptibility, etc) is needed
before the existence of a phase transition at $T^*$ may become accepted. It is not
our aim to describe the two transitions in any detail. We merely emphasize  that,
given the symmetry of URu$_2$Si$_2$, octupolar ordering can be followed by quadrupolar
ordering; but then further induced magnetic multipoles should be observable.

While circumstantial evidence for octupolar ordering looks encouraging, attempts
for its direct verification have as yet yielded negative results. An early,
very specific neutron scattering investigation ruled out either $B_{1u}$ or
$B_{2u}$ octupoles at the selected  wavevectors, including the (0,0,1) periodicity
of the weak-moment antiferromagnetism\cite{walker}. It has been argued that
resonant X-ray scattering would not see the octupoles\cite{nagao}.
Last, but not least, though the octupolar scenario is as yet the only one to account
for the observed fact that transverse uniaxial stress induces $\parallel{\hat z}$
antiferromagnetism, in a simple mean field version\cite{kf} it forces a choice
between $\sigma\parallel(100)$ and $\sigma\parallel(110)$ directions of stress, while
experiments seem to tell us that these directions are equivalent\cite{yokoyama}.
An essential insight is missing here.

A microscopic theory will have to address the questions raised above. Our present
investigation is of a limited scope: we use general  arguments to classify
the symmetry-allowed ordering transitions of URu$_2$Si$_2$. We are particularly
interested in the possibility of {\sl a sequence of such continuous phase transitions}.
We have no statement to make if the lower-temperature transition is of first order.

We address the simplest questions for which no knowledge of microscopic details is
required: Once we had an octupolar ordering transition, is there any compelling
reason to expect a second symmetry-breaking transition? What may be the order
parameters?  Should we expect still more, as yet undiscovered, phase transitions?

In an abstract sense, our question is the following: assuming the symmetry of
the high-temperature ("para") symmetry group ${\cal G}_{\rm para}$ has been
broken by a spontaneous ordering transition, introducing order
$\langle{\cal O}_1\rangle\ne 0$ lowers the symmetry to
${\cal G}({\cal O}_1) \in {\cal G}_{\rm para}$. What is the structure of
${\cal G}({\cal O}_1)$? What is the new classification of the order parameters?
Assuming that such order parameters are found, does this imply that further
continuous phase transitions necessarily happen?

In Ref.\cite{kf}, we discussed in some detail two questions of this nature:
the symmetry classification of the order parameters in the presence of an 1)
external magnetic field ${\cal B}\parallel{\hat z}$, and 2) a uniaxial stress.
Formally the same question arises if instead of externally applied fields,
we assume the presence of an effective field associated with either  a dipole
ordering transition (1) or quadrupolar order (2).

In  Sec.~\ref{sec:dipole}, we rephrase our earlier results on the effect of
an external magnetic field, and add some remarks about  the field  direction
dependence. In  Sec.~\ref{sec:octupole}, we analyze the symmetry in the presence
of a ${\cal T}_{xyz}$ octupolar (effective) field, and the possibility of a
sequence of phase transitions.  Sec.~\ref{sec:mfield} illustrates the general
arguments with a simple mean field calculation.

\section{Symmetry lowering in external, or effective, fields}

\subsection{Magnetic field}\label{sec:dipole}

In the presence of an external magnetic field $H_z\parallel (0,0,1)$ or equivalently
$\langle J_z\rangle >0$, the remaining purely geometrical symmetry elements are
${\cal E}$, $2{\cal C}_4$, and ${\cal C}_2^2$ (and, naturally, the inversion
${\cal I}$). The geometrical symmetry (described by unitary operations) is
lowered to ${\cal C}_{4h}$, but the
full\footnote{Magnetic field is invariant under space inversion. It is understood
that all symmetry groups would get doubled if we included the inversion  ${\cal I}$.}
symmetry group ${\cal G}_{\rm tetr}(J_z)$ contains also non-unitary elements,
namely  $2{\cal C}_2^{\prime}{\cal T}$, and $2{\cal C}_2^{\prime\prime}{\cal T}$,
where ${\cal T}$ is time reversal. ${\cal G}_{\rm tetr}(J_z)$ contains eight elements,
its character table (and multiplication table) is the same as that of the
tetragonal point group ${\cal D}_4$. The character table is given in Table~\ref{tab:jz}.

Reducing the symmetry from ${\cal G}_{\rm tetr}$ to ${\cal G}_{\rm tetr}(J_z)$,
some of the originally inequivalent irreps of ${\cal G}_{\rm tetr}$ become equivalent.
This parentage of the irreps is shown  in Table~\ref{tab:jz}.  It also follows that the
corresponding order parameters get mixed. This was discussed in Ref.~\cite{kf}.

\begin{table}[h]
\caption{\footnotesize Character table of the non-unitary symmetry group
${\cal G}_{\rm tetr}(J_z)$ of a  tetragonal system placed in a magnetic field ${\bf B}=(0,0,B_z)$.
Space inversion is omitted.}
\centering\vspace{8pt}
\begin{tabular}{c|ccccc|c|c|}
${\rm irrep}$ & ${\cal E}$ & $2{\cal C}_4$ & ${\cal C}_4^2$ &
$2{\cal C}_2^{\prime}{\cal T}$ & $2{\cal C}_2^{\prime\prime}{\cal T}$
 & parentage & operators \\
\hline\hline
 $A_{1}$ & 1 & 1 & 1 & 1 & 1  & $A_{1g}$, $A_{2u}$ & {\bf 1}, $J_z$ \\
 $A_{2}$ & 1 & 1 & 1 & -1 & -1 &  $A_{1u}$, $A_{2g}$  & ${\overline{J_xJ_y(J_x^2-J_y^2)}}$,
  ${\overline{J_xJ_yJ_z(J_x^2-J_y^2)}}$ \\
 $B_{1}$ & 1 & -1 & 1 & 1 & -1 &  $B_{1g}$, $B_{2u}$ & ${\cal O}_2^2$,
 ${\cal T}^{\beta}_z={\overline{J_z(J_x^2-J_y^2)}}$ \\
 $B_{2}$ & 1 & -1 & 1 & -1 & 1  &  $B_{2g}$, $B_{1u}$ & ${\cal O}_{xy}$, ${\cal T}_{xyz}$ \\
 $E$ & 2 & 0 & -2 & 0 & 0  & $E_g$, $E_u$ & \{ $J_x$, $J_y$ \} , \{ ${\cal O}_{xz}$,
 ${\cal O}_{yz}$ \} \\
  \hline\hline
 \end{tabular}
 \label{tab:jz}
\end{table}

${\hat x}$ is a lower-symmetry direction than ${\hat z}$ and correspondingly  a system
subjected to a magnetic field ${\bf B}=(B_x,0,0)$ has a smaller symmetry group
${\cal G}_{\rm tetr}(J_x)$ (Table~\ref{tab:jx}). In contrast to the case ${\bf B}\parallel{\hat z}$
where irreps taken from the zero field case kept their dimensionality only pairwise merged,
now the two-dimensional irreps split. The parentage of the ${\bf B}\parallel{\hat x}$
irreps is as follows $A_{1g}, B_{1g} \longrightarrow
\Gamma_1$, $A_{2g}, B_{2g} \longrightarrow \Gamma_2$, $A_{1u}, B_{1u} \longrightarrow \Gamma_3$,
$A_{2u}, B_{2u} \longrightarrow
\Gamma_4$, $E_g \longrightarrow \Gamma_3 +  \Gamma_4$, $E_u \longrightarrow \Gamma_1 +  \Gamma_2$.

\begin{table}[h]
\caption{\footnotesize Character table of the non-unitary symmetry group ${\cal G}_{\rm tetr}(J_x)$}
\centering\vspace{8pt}
\begin{tabular}{c|cccc}
  ${\rm irrep}$ & ${\cal E}$ & ${\cal C}_4^2{\cal T}$ &
  ${\cal C}_{2x}^{\prime}$ & ${\cal C}_{2y}^{\prime}{\cal T}$  \\
  \hline\hline
  $\Gamma_1$ & 1 & 1 & 1 & 1 \\
   $\Gamma_2$ & 1 & 1 & -1 & -1 \\
 $\Gamma_3$ & 1 & -1 & 1 & -1 \\
 $\Gamma_4$ & 1 & -1 & -1 & 1 \\
 \end{tabular}
 \label{tab:jx}

\end{table}

The presence of non-identity irreps in Tables \ref{tab:jz} and \ref{tab:jx} shows
that for fields ${\bf B}\parallel{\hat z}$, and even for ${\bf B}\parallel{\hat x}$,
a variety of symmetry breaking transitions are possible. Since the field has  broken time
reversal invariance, and induced a polarization in its direction\footnote{Assuming
the local Hilbert space allows a Zeeman splitting for the particular field direction.
E.g., the quasidoublet $\{|t_1\rangle , |t_4\rangle\}$  of URu$_2$Si$_2$ would not
split in a field ${\bf B}\parallel{\hat z}$.}, all  these transitions have the
character of orbital ordering (lifting some residual non-Kramers degeneracy), and
also lead to the appearance of time-reversal-odd transverse polarization components.
We enumerate the possibilities (${\bf B}\parallel{\hat z}$)

\begin{itemize}
\item{A field  ${\bf B}\parallel{\hat z}$ allows the ordering of $E$ type quadrupoles
and this is accompanied by the appearance of dipole polarization in the $xy$ plane.
This was suggested\cite{kf} for the disjoint high-field phase observed in
experiments\cite{jaime} on URu$_2$Si$_2$. Analogous phenomena are observed\cite{kohgi}
in high field experiments on the tetrahedral skutterudite PrOs$_4$Sb$_{12}$.}
  
\item{The ordering is of $B_1$ type. In our suggestion\cite{kf} for the low-field
 order of URu$_2$Si$_2$, the primary order was ${\cal T}_{z}^{\beta}$ octupolar
  (this survives when the field is switched off), and ${\cal O}_2^2$ quadrupoles
   were field-induced. Santini's scenario\cite{santini1} was the opposite: primary
    quadrupolar order, and field-induced octupoles.}
\item{$B_2$ order parameters:  primary ${\cal T}_{xyz}$ octupolar order and field-induced
 ${\cal O}_{xy}$ quadrupoles. In the present paper, we assume this is the hidden order
  with onset temperature $T_0=17$K. Lacking a microscopic model, it is impossible
   to decide between this scenario and the previous $B_1$ scheme.}
\item{The exotic possibility of $A_2$ order: ${\cal H}_1$ hexadecapoles and
${\overline{J_xJ_yJ_z(J_x^2-J_y^2)}}$ triakontadipoles, has not been considered yet.}
\end{itemize}

The discussion of the cases ${\bf B}\parallel{\hat y}$ and
${\bf B}\parallel({\hat x}\pm{\hat y})$ would be analogous to
${\bf B}\parallel{\hat x}$: the remaining symmetry group has order 4 (or 8, if we
 include inversion). The symmetry group for fields lying in the
$x$--$y$ plane in non-special directions is generated by ${\cal C}_4^2{\cal T}$
and ${\cal I}$. For fields in general out-of-plane directions, the only remaining
symmetry is inversion.

We conclude that a spontaneous symmetry breaking transition in external fields remains
possible if the field is either parallel or perpendicular to ${\hat z}$.  In particular,
the octupolar transition can remain second order for a range of field strengths.
However,
${\cal T}_{xyz}$ octupoles get mixed with ${\cal O}_{xy}$ ($B_2$) quadrupoles for
${\bf B}\parallel{\hat z}$, and with $E$-derived  quadrupoles (a suitable combination
of ${\cal O}_{yz}$ and ${\cal O}_{xz}$) for ${\bf B}\perp{\hat z}$. For general
field directions, a smearing of the octupolar transition should be observed. This
effect  may be helpful in deciding whether the hidden order is of octupolar nature.

The above reasoning can also  be extended to discuss further symmetry breakings
after a phase transition to a dipole-ordered phase has taken place. The effective
field of the ordered moments acts the same way as an external field. E.g., we may
conclude that following a transition to a $\langle J_z\rangle \ne 0$ phase at the
higher critical temperature $T_c^>$, tranverse dipoles  (from the $E$ doublet
$\{ J_x, J_y \}$) may also order at a lower critical  temperature $T_c^<$, and
that $\langle J_x\rangle \ne 0$ induces $\langle {\cal O}_{xz}\rangle \ne 0$. Though
symmetry allows this (or some other scenario identifiable from Table~\ref{tab:jz})
to happen, whether this potentiality is realized depends on the nature of the microscopic
model.

\subsection{The symmetry group of the octupolar phase}
\label{sec:octupole}

We may ask whether following the onset of, say,
${\cal T}_{xyz}$ type octupolar order a further symmetry lowering transition is
possible. This is the same question as to whether spontaneous symmetry breaking in
an external octupolar field of $B_{1u}$ symmetry is possible. The question is partly
academical, but it is also motivated by the NQR findings  on URu$_2$Si$_2$ by Saitoh
et al\cite{takagi}. In particular, they claimed that  either ${\cal O}_{xy}$ or
${\cal O}_2^2$ quadrupoles appear at $T^*\approx 13.5{\rm K}$, well below
$T_0\approx 17{\rm K}$ which we associate with the onset of octupolar order. However,
as we emphasized before, the connection of our arguments with experiments remains tenuous.

\begin{figure}[ht]
\vspace{0cm}\centerline{\includegraphics[height=4.3cm] {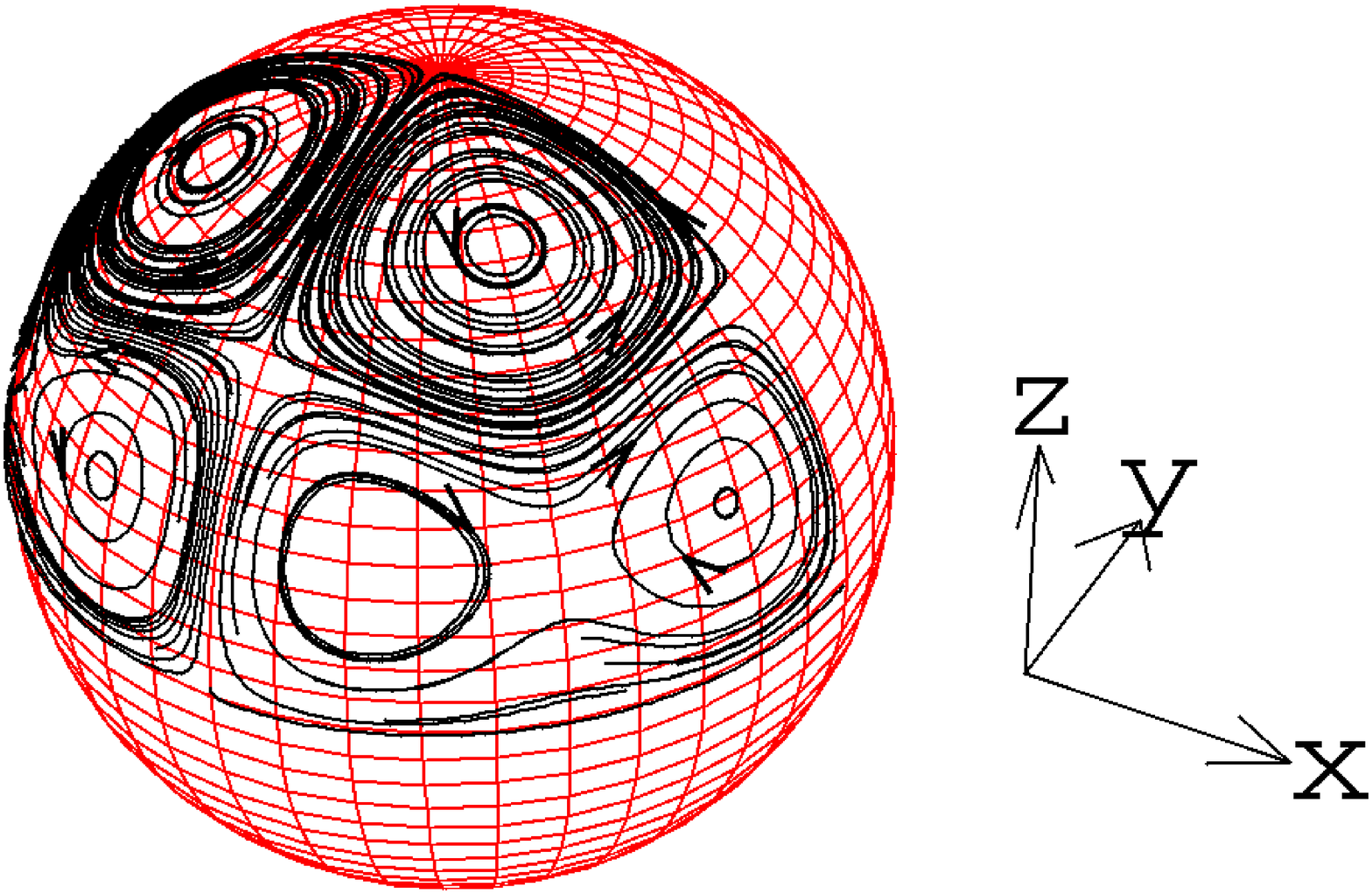}
\hspace{0cm}\includegraphics[height=4.3cm] {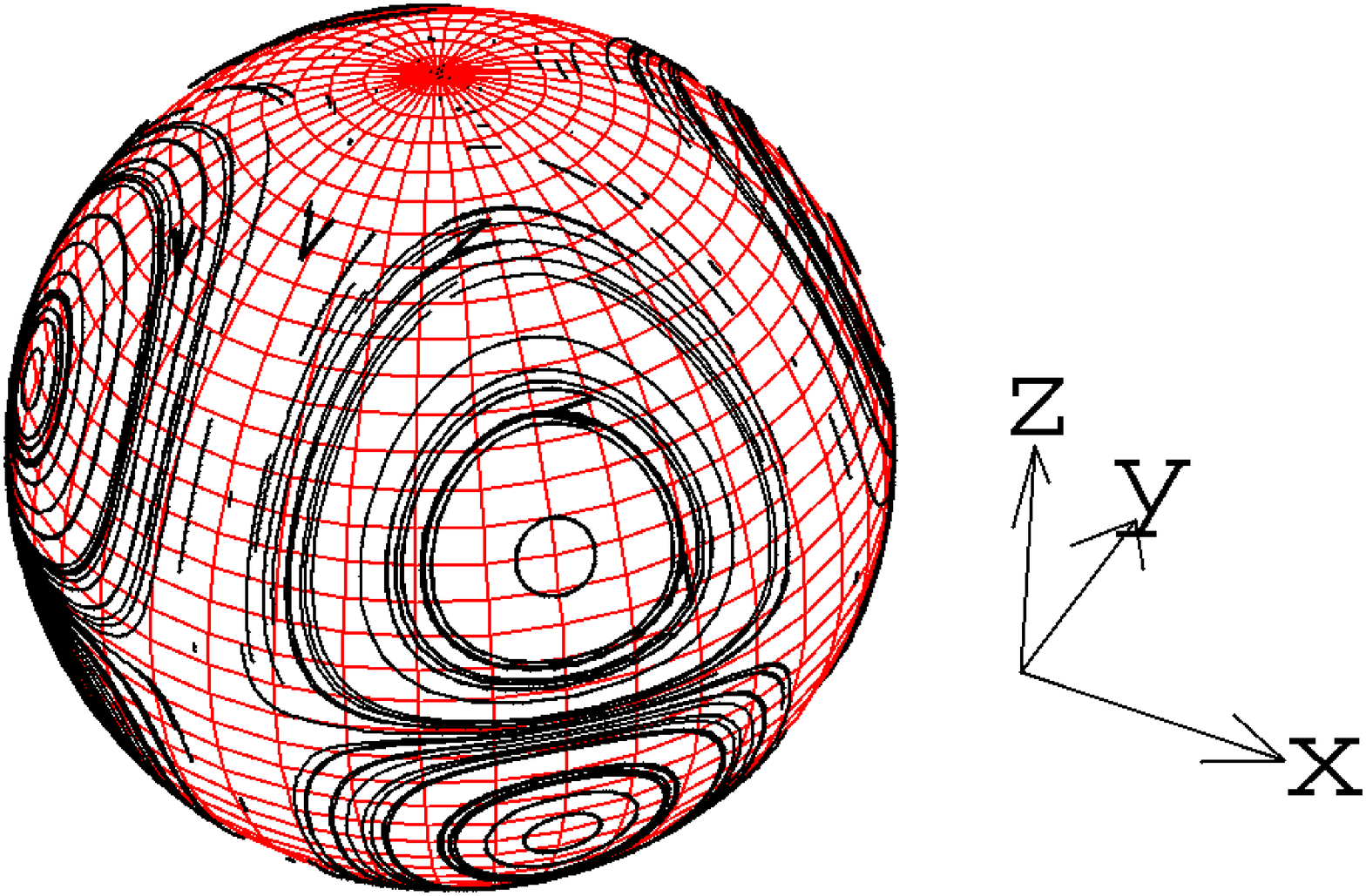}}
\vspace{0cm}\caption{Crystal field effects on the current distribution
in the ${\cal T}_{xyz}$ octupole ground state.  Left: the doublet $\{|t_1\rangle,|t_3\rangle\}$
($|t_3\rangle = (1/\sqrt{2})(|2\rangle + |-2\rangle)$). Right: the triplet
$\{|t_1\rangle, |t_2\rangle, |t_4\rangle\}$  (\protect\ref{eq:t1})--(\protect\ref{eq:t4}).
The two current distributions have the same symmetry given in
Table~\protect{\ref{tab:Txyz}}.} \label{fig:t1t3}
\end{figure}

We have to identify which operations leave the ${\cal T}_{xyz}$ octupoles unchanged,
i.e., we are looking for the octupolar symmetry group  ${\cal G}_{\rm tetr}({\cal T}_{xyz})$
as a subgroup of ${\cal G}_{\rm tetr}$. Schematically, ${\cal T}_{xyz}$  is represented in
Fig.~\ref{fig:t1t3} where the current distribution in the ground state of ${\cal T}_{xyz}$
is shown for two- and three-dimensional subspaces selected from the 9-dimensional
Hilbert space of the tetragonal crystal field eigenstates\cite{kf,santini1}. It is
obvious that the selection of the basis functions (which is "done" by the crystal
field potential) has a strong influence on the details of the current pattern. However,
though the pattern is much more decorated for  the $\{ |t_1\rangle , |t_3\rangle\}$
doublet than for the triplet $\{ |t_1\rangle , |t_2\rangle , |t_4\rangle\}$  (the
crystal field model used in Ref.\cite{kf}), the symmetries are the same. In either
case, there is a current distribution with eight major eddies: four positive, and
four negative (whether within these, there are sub-eddies, has no influence on the
symmetry). Though there are local magnetic fields, the total magnetic moment
is zero.  A ${\cal C}_4$ rotation takes positive eddies into negative ones, and
vice versa; however, reversing also the direction of currents, the original state
is restored. From the ${\cal C}_2^{\prime}$ and  ${\cal C}_2^{\prime\prime}$ elements
of ${\cal G}_{\rm tetr}$, the latter two have to be combined with ${\cal T}$. The
character table of the symmetry group ${\cal G}_{\rm tetr}({\cal T}_{xyz})$ is shown
in Table~\ref{tab:Txyz} (as before, it is understood that the space inversion
${\cal I}$ would generate the other half of the complete symmetry group). We have
also given the resulting symmetry classification of some of the order parameters
in the last column of
Table~\ref{tab:Txyz}.

\begin{table}[h]
\caption{\footnotesize Character table of the non-unitary symmetry group
${\cal G}_{\rm tetr}({\cal T}_{xyz})$ of a  tetragonal system placed in
an effective octupolar field.  Space inversion is omitted.}
\centering\vspace{8pt}
\begin{tabular}{c|ccccc|c|c|}
${\rm irrep}$ & ${\cal E}$ & $2{\cal C}_4{\cal T}$ & ${\cal C}_4^2$ &
$2{\cal C}_2^{\prime}$ & $2{\cal C}_2^{\prime\prime}{\cal T}$
 & parentage & operators \\
\hline\hline
 $A_{1}$ & 1 & 1 & 1 & 1 & 1  & $A_{1g}$, $B_{1u}$ & {\bf 1}, ${\cal T}_{xyz}$ \\
 $A_{2}$ & 1 & 1 & 1 & -1 & -1 &  $A_{2g}$, $B_{2u}$  & ${\overline{J_xJ_y(J_x^2-J_y^2)}}$,
 ${\cal T}^{\beta}_z={\overline{J_z(J_x^2-J_y^2)}}$  \\
 $B_{1}$ & 1 & -1 & 1 & 1 & -1 &  $B_{1g}$, $A_{1u}$ & ${\cal O}_2^2$,
 ${\overline{J_xJ_yJ_z(J_x^2-J_y^2)}}$   \\
 $B_{2}$ & 1 & -1 & 1 & -1 & 1  &  $B_{2g}$, $A_{2u}$ & ${\cal O}_{xy}$, $J_z$ \\
 $E$ & 2 & 0 & -2 & 0 & 0  & $E_g$, $E_u$ & \{ $J_x$, $J_y$ \} , \{ ${\cal O}_{xz}$, ${\cal O}_{yz}$ \} \\
  \hline\hline
 \end{tabular}
 \label{tab:Txyz}
\end{table}

It is interesting to note the similarities and dissimilarities to the symmetry group
of $J_z$ (Table~\ref{tab:jz}). The character table is the same only in
${\cal G}_{\rm tetr}({\cal T}_{xyz})$,  $2{\cal C}_4$ and $2{\cal C}_2^{\prime}$,
while in ${\cal G}_{\rm tetr}(J_z)$, $2{\cal C}_2^{\prime}$ and
$2{\cal C}_2^{\prime\prime}$ have to be combined with ${\cal T}$.

Analogous results would have been obtained if we had assumed a
${\cal T}^{\beta}_z$ octupolar field instead of  ${\cal T}_{xyz}$. That was
our starting assumption in Ref. \cite{kf}.

${\cal G}_{\rm tetr}({\cal T}_{xyz})$ has two generators. There are several choices:
\begin{itemize}
\item{${\cal C}_4{\cal T}$ and ${\cal C}_{2x}^{\prime}$ or ${\cal C}_{2y}^{\prime}$}
\item{${\cal C}_4{\cal T}$ and ${\cal C}_{2,x+y}^{\prime\prime}{\cal T}$ or ${\cal C}_{2,x-y}^{\prime\prime}{\cal T}$}
\item{ either of ${\cal C}_{2}^{\prime}$ and either of ${\cal C}_{2}^{\prime\prime}$}
\end{itemize}

At $T^*<T_0$, one of the order parameters appearing in Table~\ref{tab:Txyz} acquires
non-zero expectation value, and the symmetry of the system is lowered to one of the
subgroups of ${\cal G}_{\rm tetr}({\cal T}_{xyz})$. The list of the subgroups is
\begin{eqnarray}
{\cal G}_1 & = & \{ {\cal E}, {\cal C}_4{\cal T}, {\cal C}_4^2, {\cal C}_4^3{\cal T} \} \nonumber \\
{\cal G}_2 & = & \{ {\cal E},  {\cal C}_{2,x}^{\prime} \} \nonumber \\
{\cal G}_3 & = & \{ {\cal E},  {\cal C}_{2,y}^{\prime} \} \nonumber \\
{\cal G}_4 & = & \{ {\cal E},  {\cal C}_{2,x+y}^{\prime\prime}{\cal T} \} \nonumber\\
{\cal G}_5 & = & \{ {\cal E},  {\cal C}_{2,x-y}^{\prime\prime}{\cal T} \} \nonumber\\
{\cal G}_6 & = & \{ {\cal E},  {\cal C}_{2,x}^{\prime}, {\cal C}_{2,y}^{\prime},
 {\cal C}_4^2 \} \nonumber \\
{\cal G}_7 & = & \{ {\cal E},  {\cal C}_{2,x+y}^{\prime\prime}{\cal T},
{\cal C}_{2,x-y}^{\prime\prime}{\cal T},  {\cal C}_4^2 \} \nonumber \\
{\cal G}_8 & = & \{ {\cal E},   {\cal C}_4^2 \}\nonumber
\end{eqnarray}

Each of the order parameters appearing in Table~\ref{tab:Txyz} breaks one, or several,
of the symmetries in ${\cal G}_{\rm tetr}({\cal T}_{xyz})$, and thus reduces the
symmetry to one of the subgroups. All the possibilities are listed below
\begin{eqnarray}
{\cal T}^{\beta}_z {\mbox{\  \ and\ \ }}  {\overline{J_xJ_y(J_x^2-J_y^2)}}
 & \longrightarrow {\cal G}_1 \nonumber\\
{\cal O}_{yz} {\mbox{\  \ and\ \ }}  J_x & \longrightarrow {\cal G}_2 \nonumber\\
{\cal O}_{xz} {\mbox{\  \ and\ \ }}  J_y & \longrightarrow {\cal G}_3 \label{eq:opciok}\\
{\cal O}^{2}_2 {\mbox{\  \ and\ \ }}  {\overline{J_xJ_yJ_z(J_x^2-J_y^2)}}
 & \longrightarrow {\cal G}_6 \nonumber\\
{\cal O}_{xy} {\mbox{\  \ and\ \ }}  J_z & \longrightarrow {\cal G}_7 \label{eq:list}
\end{eqnarray}

Time-reversal-even and time-reversal-odd order parameters appear in pairs. Since
the octupolar background
already breaks time reversal invariance, it cannot be "unbroken" by the next
transition, so the new phase has a new pattern of non-zero current densities.
All the quadrupoles allowed by tetragonal symmetry can appear in a symmetry-breaking
transition but they bring either dipoles or higher magnetic multipoles with them.
We list the possibilities:

\begin{itemize}
\item{Most straightforward is the case of ${\cal O}_{xy}$ quadrupolar ordering
accompanied by magnetic moments in the $z$ direction. The coupling of these three
multipoles: ${\cal T}_{xyz}$, ${\cal O}_{xy}$, and $J_z$ has been discussed from
several points of view. In Ref.~\cite{kf}, we argued that on the background of
${\cal T}_{xyz}$ octupole ordering, applying uniaxial stress $\sigma\parallel(1,1,0)$
creates ${\cal O}_{xy}$ quadrupoles, and therefore magnetic moments $J_z$, offering
an interpretation of the experimental results by Yokoyama et al\cite{yokoyama}.
The existence of a third-order invariant ${\cal T}_{xyz}{\cal O}_{xy}J_z$ implies
that if in an ordered phase ${\cal T}_{xyz}\ne 0$ then also the correlator
$\langle{\cal O}_{xy}J_z\rangle\ne 0$. The non-vanishing correlator does not
automatically give non-zero values to $\langle{\cal O}_{xy}\rangle$, and
$\langle J_z\rangle$, but at least it gives a hint that ${\cal O}_{xy}$,
and $J_z$ are likely actors in a cooperative phenomenon. Here we discuss them
as coupled order parameters of a low-temperature phase transition following a
high-temperature octupolar transition.}
\item{
${\cal O}_{xz}$ (${\cal O}_{yz}$) quadrupoles would be accompanied by the
transverse magnetization component $J_x$ ($J_y$), so the experimental verification
(or refusal) of this scenario should be straightforward.}
\item{More exotic is the possibility of ${\cal O}_2^2$ quadrupolar ordering
(one of the likely candidates according to Ref.~\cite{takagi}) which should be
accompanied by
$ {\overline{J_xJ_yJ_z(J_x^2-J_y^2)}}$  triakontadipole ordering.}
\item{Finally, the symmetry of the ${\cal T}_{xyz}$ octupolar field can be
spontaneously broken by ordering the ${\cal T}^{\beta}_z$ octupoles; this should
be accompanied by hexadecapole order of the ${\cal H}_1={\overline{J_xJ_y(J_x^2-J_y^2)}}$
type. We note that ${\cal H}_1$ hexadecapoles at the U sites can combine to
an effective quadrupole at the Ru sites, so this scenario is not necessarily in
conflict with the findings by Saitoh et al.\cite{takagi}. It is an added attraction
that the simultaneous presence of $\langle{\cal T}_{xyz}\rangle\ne 0$ and
$\langle{\cal T}^{\beta}_z\rangle\ne 0$ would allow that uniaxial press applied
either in the  $\sigma\parallel(100)$ or the  $\sigma\parallel(110)$ direction
induce $J_z\ne 0$, as observed\cite{yokoyama}. We note that the measurements of
Yokoyama et al.\cite{yokoyama} were carried out at 1.4K, well below either $T_0$
or $T^*$, so both octupolar amplitudes would be near their saturation values. We
have to admit, though, that in our scenario it would be difficult to get them equal.}
\end{itemize}

\subsection{Octupolar phase in external magnetic field}

It is of some interest to combine the previous two cases to discuss the
remaining possibilities of symmetry breaking if the established ${\cal T}_{xyz}$
octupolar order is subject to an external magnetic field.

A field ${\bf B}\parallel{\hat z}$ reduces the symmetry to the four-element
subgroup ${\cal G}_7$. There are four irreps: $A_1^{\prime}$,
$A_2^{\prime}$,$A_3^{\prime}$, and $A_4^{\prime}$. In terms of the irreps shown in
Table~\ref{tab:Txyz}, their parentage is:
\[
A_1, B_2 \longrightarrow A_1^{\prime},
\]
\[
A_2, B_1 \longrightarrow A_2^{\prime}, \]
\[
E  \longrightarrow A_3^{\prime}+A_4^{\prime}
\]
Thus now ${\cal T}_{xyz}$, $J_z$, and ${\cal O}_{xy}$ are all present in the identity
representation.

A spontaneous symmetry breaking transition is possible to a  phase with
$\langle{\cal O}_2^2\rangle\ne 0$; but then ${\cal T}^{\beta}_z$, ${\cal H}_1$,
and ${\overline{J_x J_y J_z (J_x^2 - J_y^2)}}$ are induced order parameters.

There are two more possibilities of ordering, both involving a transverse dipole
 and a quadrupole (e.g., $J_x-J_y$ and ${\cal O}_{xz}-{\cal O}_{yz}$).

A field ${\bf B}\parallel{\hat x}$ reduces the symmetry to the two-element subgroup
${\cal G}_2$. Still, there remains one symmetry element to break: it can be done
by $J_z$, and a number of associated multipoles.

To conclude this subsection: if the $T<T_0$ hidden order is octupolar, the possibilities
of remaining symmetry breaking at $T^*$ depend very much on the direction of the
applied field. It seems that the only direction offering non-trivial possibilities
is ${\bf B}\parallel{\hat z}$, where the ${\cal O}_2^2$ quadrupolar symmetry breaking
is accompanied by induced octupolar, hexadecapole, and triakontadipole moments.

\section{Mean field calculations}
\label{sec:mfield}

From the fact that the symmetry group of a model contains non-trivial elements, it
does not necessarily follow that symmetry breaking phase transitions occur until
the only remaining symmetry element is the unit operator. It is possible that the
local Hilbert space is not large enough, or it does not have the right structure,
to support a sequence of ordering transitions. An ordering transition is an inevitability
only if in its absence, the ground state is degenerate. In models of octupolar
ordering, this is usually not the case. Santini's\cite{santini1} and our\cite{kf}
crystal field model is built with three singlets, so the single ion ground state is
in any case non-degenerate. Still, symmetry breaking transitions can occur by the
induced moment mechanism if the level splittings are not too large to begin with.
After the first ordering transition - which we assume is ${\cal T}_{xyz}$ octupolar
ordering - has taken place at $T_{0}$, the symmetry is reduced to
${\cal G}_{\rm tetr}({\cal T}_{xyz})$. The ionic level scheme is formed by
the combined action of the crystal field potential and the octupolar effective
field, and the ground state is again non-degenerate.  Nevertheless, the induced
moment mechanism can be effective again and we may ask whether  another second
order transition may take place at $T^*<T_{0}$, corresponding to one of the
options listed in (\ref{eq:opciok}). A mean field calculation shows that this is
indeed possible for a wide range of parameters. Naturally, one has to assume
non-zero couplings for the multipoles which appear in the definition of the
new order (for instance,  ${\cal O}^{2}_2$ and   ${\overline{J_xJ_yJ_z(J_x^2-J_y^2)}}$,
or ${\cal O}_{xy}$ and $J_z$),
but this is in any case more plausible than setting the said couplings to zero.
The starting values of the crystal field splittings do not play any particular
role, except that they have to be small enough to allow ordering.

Let us observe that ${\cal G}_{\rm tetr}({\cal T}_{xyz})$ has a two-dimensional
($E$) representation. It shows that if the crystal field scheme includes a
low-lying doublet, then this feature may be preserved also in the octupole-ordered
phase.  The two-fold degeneracy may then be lifted at the second ordering transition.
In this sense, crystal field schemes with either a doublet ground state, or a
low-lying doublet, offer a more direct route to a second symmetry breaking transition.
We note that the crystal field level scheme is not undisputed: while models with
low-lying singlets have been most widely discussed, doublet ground state was
also considered \cite{ohkawa}. However, a sequence of two continuous transitions
is possible either with, or without, a low-lying doublet.

\subsection{Three singlets}

Since $f^2$ is a non-Kramers configuration, it is possible to choose time reversal
invariant basis functions. For the singlets  $A_1$, $A_2$, and $B_2$,  these are
\begin{eqnarray}
|t_1\rangle & = & b\left( |4\rangle + |-4\rangle \right) + \sqrt{1-2b^2}
|0\rangle \label{eq:t1}\\
|t_2\rangle & = & \frac{i}{\sqrt{2}} \left( |4\rangle - |-4\rangle \right) \label{eq:t2} \\
|t_4\rangle & = & \frac{i}{\sqrt{2}} \left( |2\rangle - |-2\rangle \right) \label{eq:t4}
\end{eqnarray}
In this basis, electric (magnetic) multipoles are real (imaginary) operators (see Appendix).

Choosing the subspace of the singlets (\ref{eq:t1}-\ref{eq:t4}) is the common starting
point for a mean field calculation in Santini's\cite{santini1} and our\cite{kf} work.
The presence of these three states is useful for getting good overall agreement with
measured macroscopic quantities.
Depending on the assumptions about the sequence and the splitting of the levels, and
about the matrix elements connecting them, the three-singlet scheme may support $J_z$
dipolar, ${\cal O}_2^2$ or ${\cal O}_{xy}$ quadrupolar, ${\cal T}^{\beta}_z$ or
${\cal T}_{xyz}$ octupolar, or ${\cal H}_1$ hexadecapole order in the first ordered phase,
and a combination of two more orders from the same list below  a second critical
temperature. Up to now, mainly scenarios with a single transition were considered.
Santini\cite{santini1} chose quadrupolar order while we preferred octupolar order.
There is no effective way to choose between
${\cal T}^{\beta}_z$ and ${\cal T}_{xyz}$. Our earlier discussion\cite{kf} was based
on postulating  ${\cal T}^{\beta}_z$ ordering. Here we phrase our arguments on the
alternative assumption that ${\cal T}_{xyz}$ order appears first.

The energy (free energy) density can be expressed as a sum of invariants. A number of
third order invariants contain ${\cal T}_{xyz}$. The corresponding third-order
contribution to the free energy is
\begin{eqnarray}
\Delta F_3 & = & c_1({\bf Q}_1,{\bf Q}_2,{\bf Q}_3){\cal T}_{xyz}({\bf Q}_1)
{\cal T}_{\beta}^z({\bf Q}_2){\cal H}_1({\bf Q}_3) \nonumber\\
& & +
c_2({\bf Q}_1,{\bf Q}_2,{\bf Q}_3){\cal T}_{xyz}({\bf Q}_1){\cal O}_{xy}
({\bf Q}_2)J_z({\bf Q}_3) \nonumber\\
& & + c_3({\bf Q}_1,{\bf Q}_2,{\bf Q}_3){\cal T}_{xyz}({\bf Q}_1)
\left[ J_x({\bf Q}_2){\cal O}_{zx}({\bf Q}_3)+J_y({\bf Q}_2){\cal O}_{yz}({\bf Q}_3)\right] \nonumber\\
 & & + c_4({\bf Q}_1,{\bf Q}_2,{\bf Q}_3){\cal T}_{xyz}({\bf Q}_1){\cal O}_2^2({\bf Q}_2)
 {\cal K}({\bf Q}_3)\label{eq:third}
\end{eqnarray}
where ${\bf Q}_1+{\bf Q}_2+{\bf Q}_3={\bf 0}$, and ${\cal K}={\overline{J_xJ_yJ_z(J_x^2-J_y^2)}}$.

If ${\cal T}_{xyz}({\bf Q}_1)$ is the hidden order, i.e.,
$\langle{\cal T}_{xyz}({\bf Q}_1)\rangle \ne 0$ at $T<T_0$, then in the same temperature
range the double correlators read off from above become non-zero, e.g.,
$\langle{\cal O}_{xy}({\bf Q}_2)J_z({\bf Q}_3)\rangle\ne 0$. This is helpful, but in
itself not sufficient, for the ordering of ${\cal O}_{xy}$ and $J_z$ separately.
For that, dipole-dipole and quadrupole-quadrupole interactions have to be non-zero,
and sufficiently strong to overcome the level splittings arising from the combined
effect of the crystal field and the octupolar effective field. Playing with  parameters
introduces a lot of arbitrariness into a crystal field model, and at present it
would be  pointless to try to fit them to experiments, especially as there is no
agreement about the relevant {\bf Q} vectors. We merely wish to  demonstrate the
possibility of a second ordering transition at $T^*<T_0$.

\begin{figure}[ht]
 \vspace{-.1cm}
 \centerline{\hspace{1cm}
 \includegraphics[height=6.5cm]{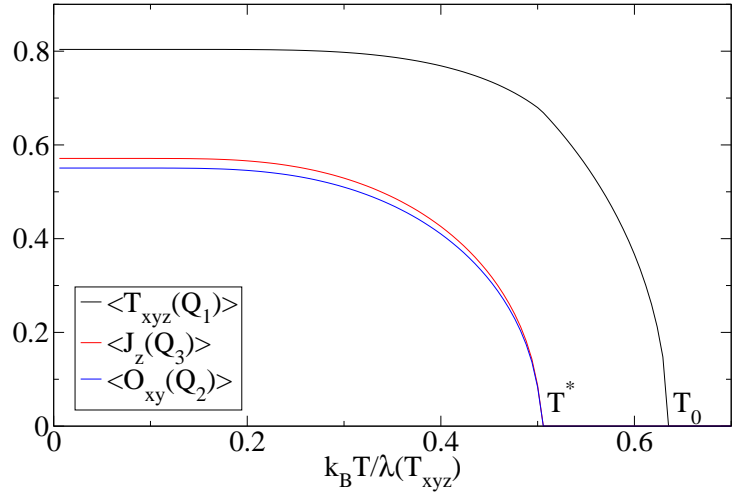}}
  \vspace{-.1cm}
  \caption{Mean field solution for two consecutive phase transitions.
  Following the onset of staggered ${\cal T}_{xyz}$ octupolar order,
  the coupled quadrupolar ${\cal O}_{xy}({\bf Q}_2)$ and dipolar
  $J_z({\bf Q}_3)$ order parameters appear at a lower critical temperature
   (the parameters are given in the text. Note the break in the slope of
   $\langle{\cal T}_{xyz}({\bf Q}_1)\rangle$ at $T=T^*$. This anomaly will
   be clearly seen in Fig.~\protect\ref{fig:anomalies1} (right).} \label{fig:meanf}
 \end{figure}

For the order parameters appearing at $T^*$, any of the pairs in
(\ref{eq:list}) could be chosen. For the moment, we arbitrarily pick ${\cal O}_{xy}$ and  $J_z$.

Ordering is possible by the induced moment mechanism in spite of crystal field
splittings, but its nature is basically the same as it would be for three degenerate
singlets\footnote{Crystal field splittings  establish an asymmetry between  the
complementary variables   ${\cal O}_{xy}$ and  $J_z$ because ${\cal O}_{xy}$ has a
matrix element between $|t_1\rangle$ and $|t_2\rangle$, while $J_z$ between
$|t_1\rangle$ and $|t_4\rangle$. Asymmetry arises also from unequal multipole
couplings.}. We introduced three mean field couplings: $\lambda({\cal T}_{xyz})$,
$\lambda({\cal O}_{xy})$, and $\lambda(J_z)$, and solved the self-consistency equations
for the three order parameters.

\begin{figure}[ht]
 \vspace{0.3cm}\centerline{\hspace{-0.2cm}
 \includegraphics[height=4.2 cm] {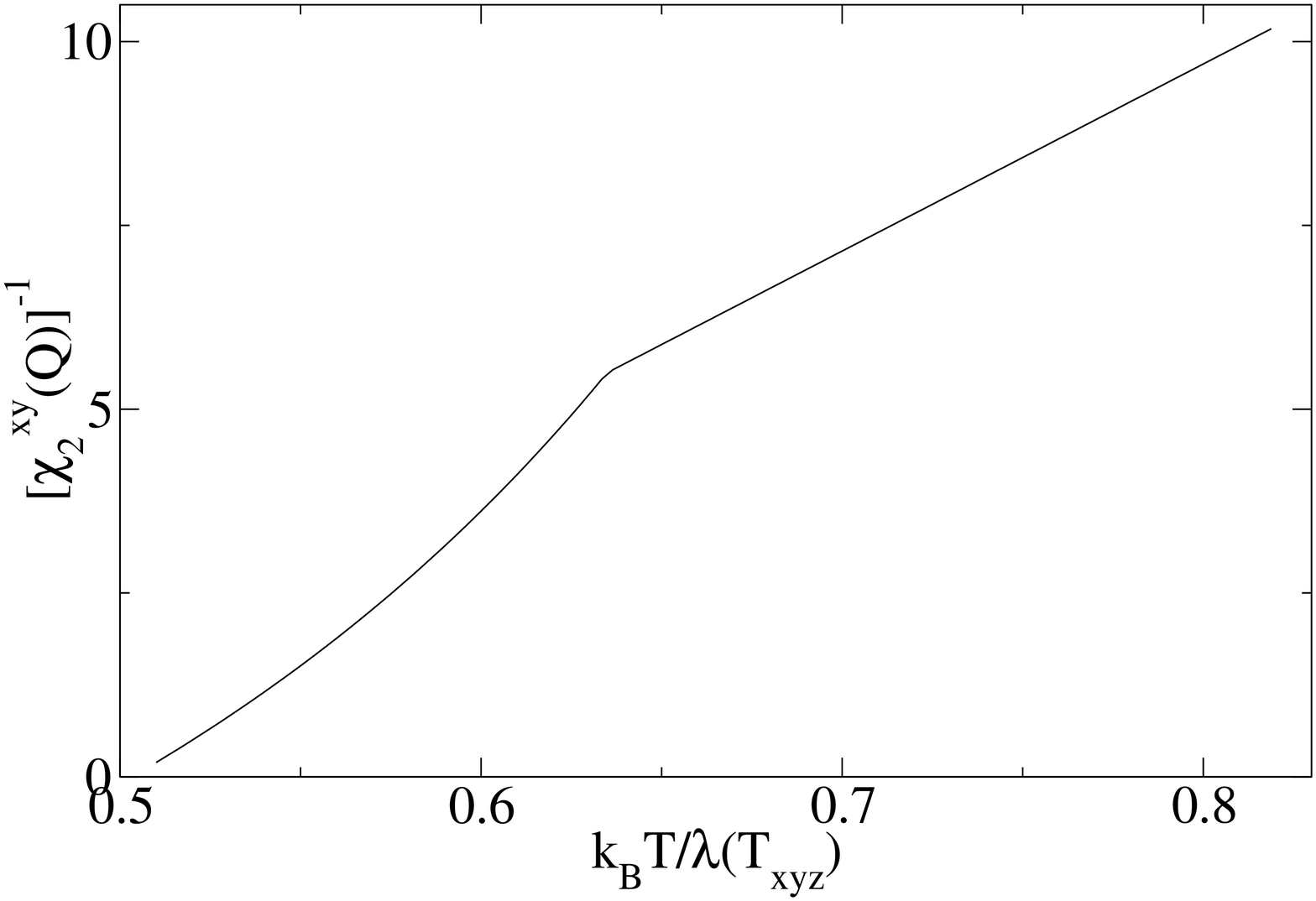}\hspace{0.2cm}
 \includegraphics[height=4.2 cm] {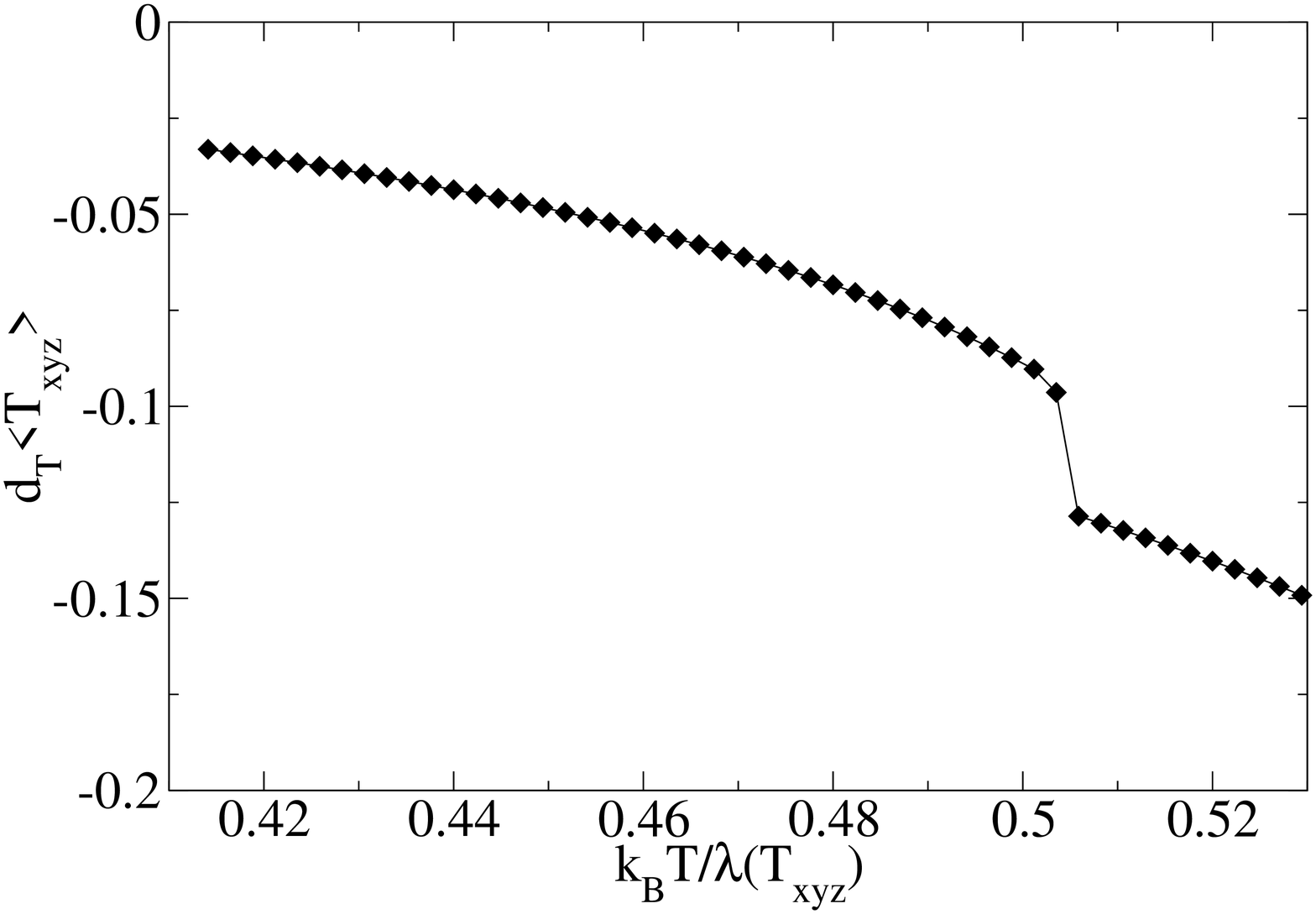}}
 \vspace{.5cm}
 \caption{Left: The temperature dependence of the inverse staggered quadrupolar susceptibility.
 Right: the discontinuity in the temperature derivative of the octupolar moment at $T^*$.}
  \label{fig:anomalies1}
 \end{figure}

Typical results are shown in Fig.~\ref{fig:meanf}.  The onset of ${\cal T}_{xyz}$ octupolar
order at $T_0=0.645$ is followed by the ordering of ${\cal O}_{xy}$ and $J_z$ at
$T^*\approx 0.495$. To bring out the contrast\footnote{The critical temperature is
the same, but the amplitudes unequal.} between the complementary parameters ${\cal O}_{xy}$
and $J_z$, we used the parameter set $\Delta_{21}=1$ as the energy unit,
$\Delta_{41}/\Delta_{21}=2$, $\lambda({\cal O}_{xy})=\lambda(J_z)$,
$\lambda(J_z)/\Delta_{21}=10$, and $\lambda({\cal T}_{xyz})/\Delta_{21}=17$.

The second step of ordering is assisted by the pre-existing octupolar order. This
can be seen in the susceptibility plot Fig.~\ref{fig:anomalies1} (left). We
have chosen the quadrupolar susceptibility which belongs to the ordering degree
of freedom ${\cal O}_{xy}({\bf Q}_2)$.  The high-temperature susceptibility extrapolates
to a lower quadrupolar ordering temperature than the behavior calculated within the
octupolar phase. An alternative argument is based on the Landau expansion (we drop
the {\bf Q}s)
\begin{eqnarray}
{\cal F} & = & a_{\rm oct}(T-T_0) {\cal T}^2 + b_{\rm oct}{\cal T}^4 +
a_{\rm quad}(T-T_1)({\cal O}^2+J^2) \nonumber\\
& & + b_{\rm quad}({\cal O}^4+J^4) +
e {\cal T}^2({\cal O}^2+J^2) +f {\cal T}{\cal O}J
\end{eqnarray}
The last term corresponds to (\ref{eq:third}). Whatever the sign of $f$, the sign of ${\cal O}J$ can be chosen so that the term lowers the free energy, and enhances the lower critical temperature from $T_1$ to $T^*$, as observed.

The octupolar order responds to the onset of quadrupolar-dipolar order
by a change of the slope of the temperature dependence (Fig.~\ref{fig:anomalies1}, right).

We emphasize that we did not make any attempt to fine-tune our mean field parameters. If
we plotted the specific heat, it would show two lambda-anomalies, in conflict with the
experiments known to us. We think that until further experiments are done, it would be
pointless to refine our calculation.

\section{Conclusion}

We studied possible sequences of symmetry breaking transitions in tetragonal $f^2$
systems. Assuming that octupolar order sets in first, the next transition may lead
to mixed dipole--quadrupole, quadrupole--triakontadipole, or octupole--hexadecapole
order. There is an interesting similarity to the recent NQR  results by Saitoh et
al\cite{takagi} but we feel that any detailed comparison would be premature.

\section*{Acknowledgments}
We are greatly indebted to S. Takagi for valuable advice, and for informing us
about his results prior to publication. We thank H. Amitsuka, F. Bourdarot, W.J.L.
Buyers, B. F{\aa}k, G. Kriza, and G. Solt for enlightening discussions and/or
correspondence. At all points of this work, we were helped by advice from,
and discussions with, K.  Penc. We were supported by the Hungarian National
Grants T038162, T049607, and TS049881.

\appendix
\section{}

The off-diagonal multipoles in basis (\ref{eq:t1}-\ref{eq:t4}) are
\begin{equation}
J_z  =  4\sqrt{2}b\left( \begin{array}{ccc}
                      0 & -i & 0 \\
                      i & 0 & 0 \\
                      0 & 0 & 0
                 \end{array}
                 \right) \\ {\mbox{\ \ \ \ \ \ }}
{\cal O}_{2}^{2}  =  2\sqrt{7}
       \left( \begin{array}{ccc}
                      0 & 0 & 0 \\
                      0 & 0 & 1 \\
                      0 & 1 & 0
                 \end{array}
                 \right) \\
\end{equation}

\begin{equation}
{\cal O}_{xy}  = (-2\sqrt{14}b + 6\sqrt{5-10b^2})
       \left( \begin{array}{ccc}
                      0 & 0 & 1 \\
                      0 & 0 & 0 \\
                      1 & 0 & 0
                 \end{array}
                 \right) \\
       \end{equation}
       \begin{equation}
{\cal T}_{z}^{\beta}  =  3\sqrt{3}(\sqrt{70}b+5\sqrt{1-2b^2})
       \left( \begin{array}{ccc}
                      0 & 0 & -i \\
                      0 & 0 & 0 \\
                      i & 0 & 0
                 \end{array}
                 \right) \\
\end{equation}

\begin{equation}
{\cal T}_{xyz}  =  3\sqrt{105}
       \left( \begin{array}{ccc}
                      0 & 0 & 0 \\
                      0 & 0 & -i \\
                      0 & i & 0
                 \end{array}
                 \right) \\ {\mbox{\ \ \ \ \ \ }}
{\cal H}_{1}  =  3\sqrt{35-70b^2}
       \left( \begin{array}{ccc}
                      0 & 1 & 0 \\
                      1 & 0 & 0 \\
                      0 & 0 & 0
                 \end{array}
                 \right)
\end{equation}

All the matrix elements of the triakontadipole operator
$ {\overline{J_x J_y J_z (J_x^2 - J_y^2)}} $ vanish in this subspace.

%



\begin{thebibliography}{99}
\bibitem{thesame} We may observe that A) is difficult to satisfy since  in Table 1.,
$J_z$ is the only operator with $A_{2u}$ symmetry. Naturally, Table 1. is
not exhaustive, and more complicated operators like plaquette dipoles, may have
the same symmetry as $J_z$.
\end{thebibliography}
\end{document}